\documentclass{ifacconf}

\usepackage{graphicx}      
\usepackage{natbib}        
\usepackage{amsmath}
\usepackage{amssymb}
 \usepackage{xcolor}
 \newcommand{\MK}[1]{{\color{black} #1}}
 \newcommand{\REV}[1]{{\color{black} #1}}
 \newcommand{\MS}[1]{{\color{black} #1}}%
\begin{document}
\begin{frontmatter}

\title{Space-Filling Input Design for Nonlinear State-Space Identification \thanksref{footnoteinfo}} 

\thanks[footnoteinfo]{Funded by the European Union (ERC, COMPLETE, 101075836). Views and opinions expressed are however those of the author(s) only and do not necessarily reflect those of the European Union or the European Research Council Executive Agency. Neither the European Union nor the granting authority can be held responsible for them.}

\author[First]{Máté Kiss}  
\author[First,Second]{Roland Tóth}
\author[First]{Maarten Schoukens}

\address[First]{Control System group, Eindhoven University of Technology,
Eindhoven, the Netherlands}
\address[Second]{Systems and Control Laboratory, Institute for Computer Science
and Control, Budapest, Hungary}

\begin{abstract}
The quality of a model resulting from (black-box) system identification is highly dependent on the quality of the data that is used during the identification procedure. Designing experiments for linear time-invariant systems is well understood and \MK{mainly focuses on} the power spectrum of the input signal. Performing experiment design for nonlinear system identification on the other hand remains an open challenge as \MK{informativity of the data}  depends both on the frequency-domain content and on the time-domain evolution of the input signal. Furthermore, as nonlinear system identification is much more sensitive to modeling and extrapolation errors, having experiments that explore the considered operation range of interest \MK{is} of high importance. Hence, this paper focuses on designing space-filling experiments \MK{i.e., experiments that cover the full operation range of interest}, for nonlinear dynamical systems that can be represented in a state-space form using a broad set of input signals. 
The presented experiment design approach can straightforwardly be extended to a wider range of system classes (e.g., NARMAX). The effectiveness of the proposed approach is illustrated \MK{on the experiment design for} a nonlinear mass-spring-damper system, using a multisine input signal.
\end{abstract}


\end{frontmatter}

\section{Introduction}
In system identification, we aim to \MK{capture} a mathematical model \MK{of the system behavior} from noisy measurement data. \MK{The success of such a process hinges on} the quality and informativity of the available data. While this is already \MK{important issue} for linear time-invariant (LTI) system identification, this becomes even more important for nonlinear system identification problems. Indeed, whereas LTI models can be thought of as a hyperplane in a high-dimensional (state) space, nonlinear models are characterized by a manifold in the same high-dimensional space \citep{Schoukens19}. In practice, this means that approximation errors are often a dominant source of errors in nonlinear system identification. Hence, nonlinear models are much more sensitive to extrapolation and modeling errors across the considered range of operation.

Historically, experiment design focused on so-called optimal experiment design, which aims to minimize the variance of the estimated parameters of the identified model, while minimizing the cost of the \MK{required} experiment. This has been studied extensively for the LTI case (e.g. \citet{Annergren2017,Bombois2021,Hjalmarsson2007}), where the final model quality largely depends on the frequency-domain energy content that is present in the input signal.

Experiment design for nonlinear system identification on the other hand is much less understood. In this case, the optimal experiment becomes dependent both on the time-domain amplitude distribution and the frequency-domain energy content. Most available results focus on rather restrictive model classes such as Hammerstein, Wiener, and similar model structures, or require the use of a specific parametrization of the input signal \citep{Cock_Phd,Cock_D_optimal,Vincent2010,Forgione2014,Colin20}.

However, as argued above, the dominant challenge in most black-box nonlinear system identification applications is not to have a small variance on the final parameter estimates, but rather to ensure that the model obtains a good data fit over the considered range of operation. Hence, some research effort has focused over the last years on space-filling experiment design. For example, \citep{Nelles2017,Nelles21,Nelles22} explored various approaches to perform space-filling experiment design while underlyingly assuming a NARMAX model structure, and \citep{Novarra2007} observed that the space-filling property resulted in more accurate model estimates in a set-membership setting. Generating space-filling data has also received attention from a data-driven (safe) model predictive control design point-of-view, for instance in \citep{Capone20}, where the authors use the MPC controller to drive the (unstable) system to regions of the operation space where no or little data is available for identification.

Contrary to the above-described NARMAX results, and motivated by \MK{the availability of powerful} nonlinear state-space system identification approaches \citep{Paduart2010,Beintema2023}, this work aims to develop a space-filling experiment design approach for \MK{state-space systems. The proposed approach is extendable for a broad class of systems such as NAR(MA)X, nonlinear output error, etc.} while allowing to use a broad class of input signals (e.g., multisine, multistep). Such a flexible parameterization of the input signal, together with the use of \MK{constrained optimization algorithm}, also allows for the inclusion of amplitude or energy constraints during the experiment design.

This paper is structured as follows. First, Section~\ref{sec:problem} introduces the considered system class and problem setting. The proposed space-filling experiment design algorithm is introduced next in Section~\ref{sec:approach}. This section \MK{represents} the considered input parametrization, the cost function, and the optimization strategy. Finally, Section~\ref{sec:example} investigates the effectiveness of the proposed approach on a nonlinear mass-spring-damper system simulation example, \MK{followed by the conclusion part in Section~\ref{sec:Conclusion}.}
\section{Problem Statement}\label{sec:problem}

\MK{Designing an optimal input is dependent on the unknown true system. However, to obtain a system model a high-quality experiment is needed. This results in the so-called \emph{chicken-and-egg problem} encountered in optimal experiment design. In this work, for simplicity, we assume that we either know the system for which we want to do the experiment design or a surrogate model (given by a white-box or black-box model) of it which we will consider as a faithful representation of the original system.}

Consider a known discrete-time system that can be represented in a state-space (SS) form:
    \begin{align}
        \begin{aligned}
            x_{k+1} &= f(x_k,u_k) \\ \label{eq:DT system}
            y_k &= g(x_k,u_k),
        \end{aligned}
    \end{align}
%
where $x_k\in\mathbb{R}^{n_{x}}$, $u_k\in\mathbb{R}^{n_{u}}$, $y_k\in\mathcal{Y}\subseteq\mathbb{R}^{n_\mathrm{y}}$ are the state, input, and output \MK{signals} at time instant \MK{$k\in\mathbb{Z}_0^+$}. The system is assumed to be BIBO stable and noiseless \MS{while functions $f$ and $g$ of \eqref{eq:DT system} are considered to be differentiable, i.e. they are part of $\mathcal{C}^1$.}

\begin{figure}[t]
    \centering
    \includegraphics[width=0.23\textwidth]{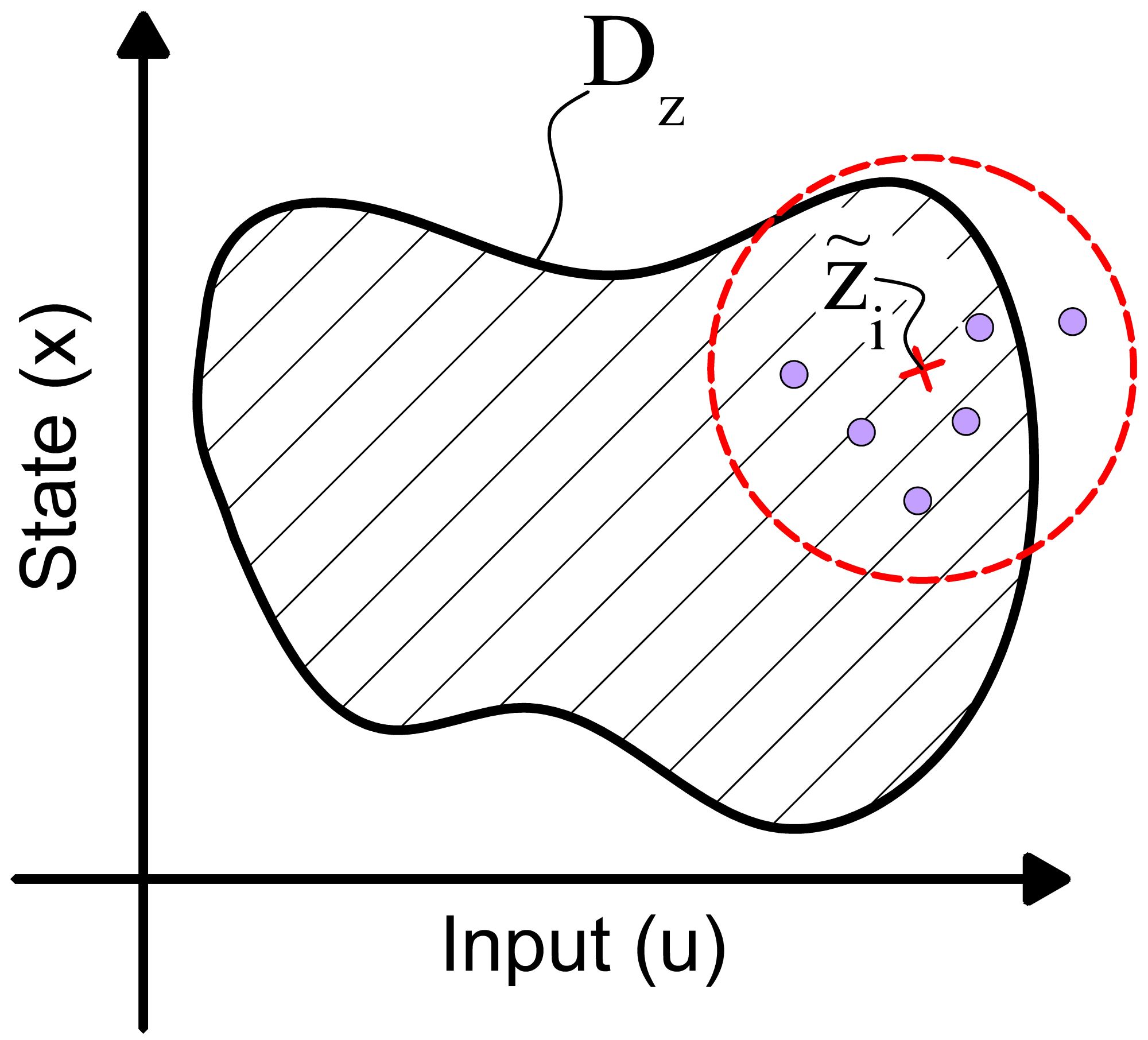}
    \caption{{\MS{Region of interest ($\mathcal{D}_z$) in the joint input-state space with a center point ($\Tilde{z}_i$). The red circle defines the local region around the center point with some data points (blue dots) in it.}}}
    \label{fig:Region of interest}
\end{figure}

Starting from the known system \eqref{eq:DT system}, this paper aims to design an input \MS{sequence $\{u_k\}_{k=0}^{N}$} such that the joint input-state samples $z_k  = \left[ u_k^\top \: x_k^\top \right]^\top$ fill a predefined domain $\mathcal{D}_z$ \MS{(Figure~\ref{fig:Region of interest})} as well as possible. \MS{A region of interest is a specific compact subset of the joint input-state space ${\mathcal{D}_z}\subseteq\mathcal{D}$, which is chosen by the user.} Note, however, that the state $x$ cannot be manipulated directly, as it is completely dependent on the input $u_k$ and its past samples through \eqref{eq:DT system}. This further challenges the space-filling objective. Note that, when considering the NAR(MA)X case, one could consider the joint space of past inputs and past outputs instead.

\section{Input design approach}\label{sec:approach}
This section outlines the proposed approach, starting from the input parametrization, followed by the proposed space-filling cost function and the utilized optimization approach.

\subsection{Input Parametrization} \label{sec:parameterizatoin}
The proposed approach allows for a broad class of input signals, we only require that the input \MS{$\{u_k\}_{k=0}^{N}$} is parametrized by $\theta$ such that the partial derivatives $\frac{\partial u_k}{\partial \theta}$ exist. 

\MK{Example of such input parametrizations include multisine signals that excites the first $F$ frequencies:}
\begin{align}\label{eq:general multisine}
\MK{u_k = \sum_{l=1}^F A_l \sin\left( 2\pi l \frac{f_0}{f_s} k + \varphi_l \right)}
\end{align}
where the parameters $\theta$ are given by the amplitude $A_l$ and/or phase \MK{$\varphi_l$} of all excited frequencies, and $f_0 = \frac{f_s}{N}$ with $f_s$ being the sampling frequency and \MK{$N\in\mathbb{Z}_0^+$} being the total number of samples per period.  Alternatively, one could also consider a direct parametrization of the input sequence $u$ of $N$ samples where each sample $u_k$ is a parameter ($u_k=\theta_k$).

\subsection{Space-Filling Cost Function} \label{section:cost}

Space-filling designs in a static setting are often obtained by solving maximin space-filling cost \citep{Nelles22}. However, optimization in such a setting is computationally costly. \MK{Hence, a different criterion is put forward here starting from a nonparametric nonlinear function estimation point of view.}

One of the most simple nonparametric nonlinear function estimation algorithms is to average the function outputs (in our case $x_{k+1}$ and $y_k$ for the state and output function respectively) over a local interval around a center point \MK{$\Tilde{z}_i$} \MS{(Figure~\ref{fig:Region of interest})} at which we would like to know the function \MK{$f(\Tilde{z}_i)$}. This is also known as a Nadaraya–Watson estimator \citep{Nadaraya1964}. Furthermore, the variance on such an averaged estimate will be inversely proportional \MK{to the number of samples $z_k$ from the data sequence z, denoted as \MS{$d(z,\Tilde{z}_i)$}, that are present in this local interval.} In simple terms, if we have no data in that local region the variance is high, therefore designing an input that minimizes the variance of such a nonparametric estimate would encourage the data to be present close to every center point \MK{$\Tilde{z}_i$}, resulting in a space-filling input design.
Hence, one could aim to minimize the following cost:
\begin{align}\label{eq:cost_integral}
    \MS{C_\theta = \int_{\mathcal{D}_z} \frac{1}{d(z,\Tilde{z})} d\Tilde{z}}.
\end{align}
Note that such a cost automatically promotes a space covering input as having extra data points in a given region of the considered domain has a rapidly diminishing return due to the inverse proportional relation \eqref{eq:cost_integral}. Hence, it is more valuable to drive the system to regions of the input-state space where no data points are available. However, this introduced heuristic cost also has multiple downsides:
\begin{itemize}
    \item It requires the calculation of an integral over the considered domain $\mathcal{D}_z$;
    \item Counting the number of points \MS{$d(z,\Tilde{z}_i)$} present in an interval around a center point \MK{$\Tilde{z}_i$} results in a non-smooth cost;
    \item If no data points are present in the considered interval, an infinite cost value results due to a division by zero.
\end{itemize}
Each of these downsides is addressed in the following paragraphs.

\textbf{From integral to sum:} Instead of considering each point of the domain as a center point \MK{$\Tilde{z}_i$}, a uniform grid of a total of $n$ center points is selected, over which the average is taken. This reduces the integral to a finite sum:
\begin{align}
    \MS{C_\theta =  \frac{1}{n} \sum_{i=1}^n \frac{1}{d(z,\Tilde{z}_i)}.}
\end{align}
Note that one could also consider a non-uniform grid distribution depending on the expected complexity of the state and output function over the considered domain (e.g. adding more grid points in regions of the domain where the functions rapidly change). However, such a non-uniform grid could emphasize regions of the considered domain more than other regions if this is not properly weighted in the cost function.

\textbf{From finite to infinite support membership functions:} As counting the number of points $d_i$ present in an interval around a center point would result in a non-smooth cost, we consider an exponential membership function instead:
\begin{equation}\label{eq:squared exponential}
    \MS{d_{i} = d(z,\Tilde{z}_i) = \sum_{k=1}^N e^{-\frac{1}{2}(\Tilde{z}_i-z_k) \Sigma^{-1} (\Tilde{z}_i-z_k)\top},}
\end{equation}
where $\Sigma$ is a diagonal matrix:
\begin{equation}\label{eq:varianceMatrix}
    \MK{\Sigma = \mathrm{diag}(\sigma_1^2, \sigma_2^2, \ldots, \sigma_{n_z}^2)}
\end{equation}
where $n_z = n_x + n_u$ \MK{is the total size of the input-state space}. \MK{Observe that if $z_k = \Tilde{z}_i$, then the exponential function $e^{-\frac{1}{2}(\Tilde{z}_i-z_k) \Sigma^{-1} (\Tilde{z}_i-z_k)^T}$ equals $1$} and that the contribution, or `added information', of a sample point decreases exponentially when it is further away from the considered grid point. The variances $\sigma_i^2$ govern how fast this decay takes place. Implicitly, such a membership function makes a smoothness assumption on the considered state and output function: i.e. it is assumed that points far away from the considered center point \MK{$\Tilde{z}_i$} still have some influence on the functions evaluated in this center point.

The values $\sigma_i^2$ need to be set by the user. It is advised to choose it large enough such that the grid points have an overlapping region to ensure that data points significantly contribute to multiple grid points at a time as this improves the cost smoothness and optimization performance. The $\sigma_i^2$ values should also not be chosen too large as that would result in a cost that is insensitive to changes in the input signal parameters.

\textbf{Adding an $\epsilon$ contribution:} by adding a small, positive, contribution in the denominator of the cost function terms, the division by zero, or division by very small numbers, is avoided:
\begin{align}
    C_\theta = \frac{1}{n} \sum_{i=1}^n \frac{1}{\epsilon + d_i}.
\end{align}

This results in the following completed expression of the considered cost:
\MK{
\begin{align} \label{eq:overall cost}
    C_\theta = \frac{1}{n} \sum_{i=1}^n \frac{1}{\epsilon + \sum_{k=1}^N e^{-\frac{1}{2}(\Tilde{z}_i-z_k) \Sigma^{-1} (\Tilde{z}_i-z_k)^T}}. 
\end{align}
}
\subsection{Optimization}
The proposed cost function \eqref{eq:overall cost} is generally nonlinear in the parameters due to a) the nonlinear dependence of the input signal on the parameters, b) the nonlinear dependence of the state on the (past) inputs, c) the nonlinear dependence of the cost on the joint input-state values. The resulting cost function is minimized using a gradient-based optimization approach. Hence, the obtained solutions only converge to the closest local optimum. The use of gradient-based optimizers requires the partial derivatives of the input w.r.t. the parameters, as well as the partial derivatives of the state and output function \eqref{eq:DT system} w.r.t. the current state and input to exist, as was assumed in the earlier sections.

In practice, the cost function~\eqref{eq:overall cost} is minimized using the Matlab function \verb+fmincon+ \MK{which} also allows additional (parameter) constraints to be added during the optimization, e.g. to limit the maximum amplitude of a signal (when using a direct parametrization $u_k=\theta_k$), or to limit the energy at each frequency (when using a multisine parametrization).

\MK{Note that as the dimensionality of the region of interest increases, the computational efficiency decreases due to the rapidly growing number of grid points that is required to cover the considered space. In many practical cases, the nonlinearities are acting on certain states rather than on the full state-space, then the dimensionality of the region of interest for which we need space-covering input design can be reduced.}

\section{Simulation example}\label{sec:example}

\begin{figure}[htb]
    \centering
    \includegraphics[width=0.4\textwidth]{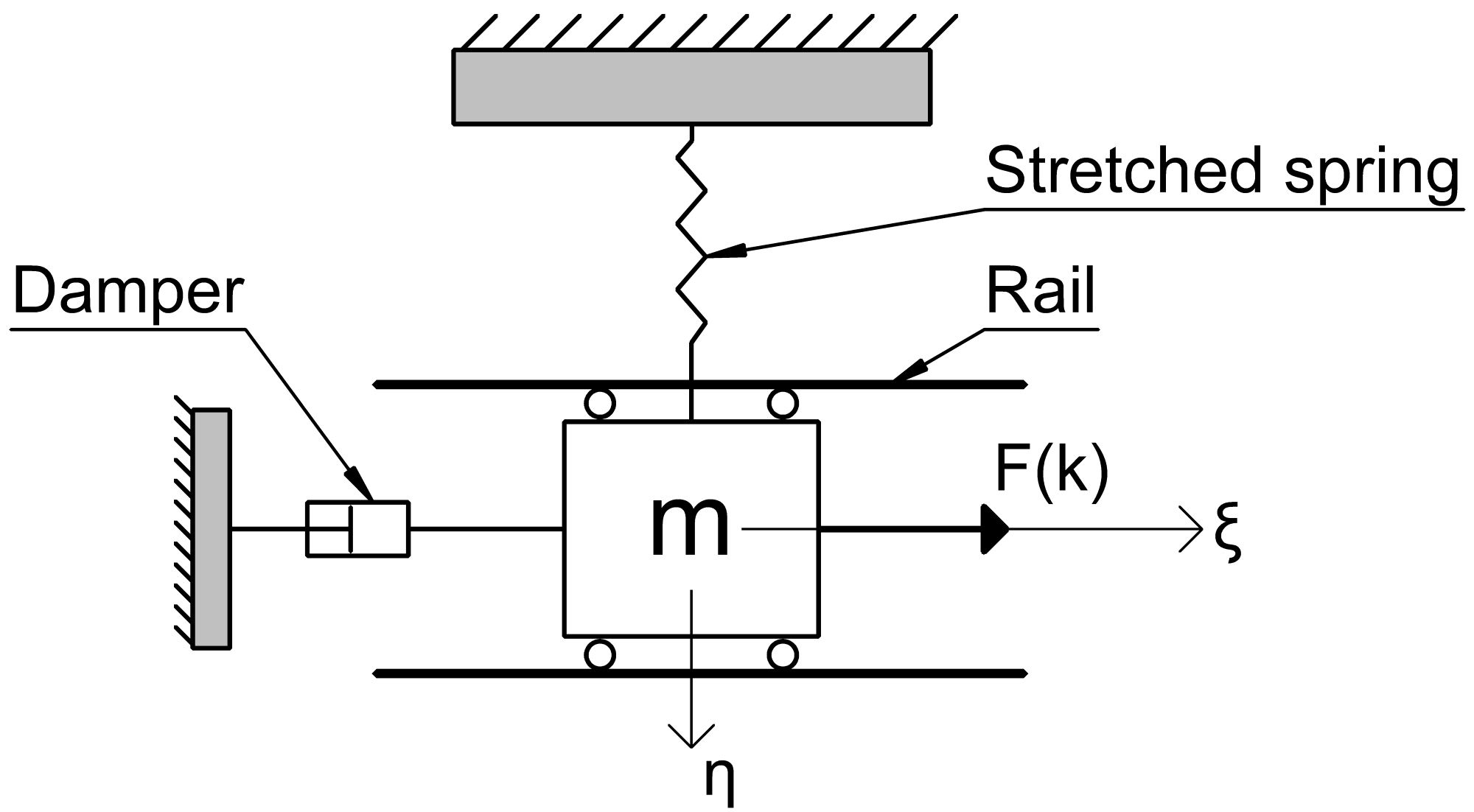}
    \caption{{Nonlinear dynamical system}}
    \label{fig:1DOF system}
\end{figure}

The effectiveness of the proposed approach is illustrated on a nonlinear 1 DOF system (Figure~\ref{fig:1DOF system}). The system consists of a horizontally moving mass fixed in a rail which is connected by a spring to the ceiling and is linearly damped. This spring connection results in a geometrical nonlinearity. The system is excited by an external force $F$.

\subsection{The considered system}
The equations of motion in a state-space form of the considered system are given by:
    \begin{align}
        \begin{gathered}
            \dot{x}_{1} = x_{2}\label{eq:1DOF system} \\
            \dot{x}_{2} = \REV{\frac{1}{m} \left(F-sx_{1} + \frac{slx_{1} }{\sqrt{x_{1}^2 + a^2}} - cx_{2}\right)},
        \end{gathered}
   \end{align}
where $x$ denotes the horizontal position of the mass ($\mbox{m=5.0 kg}$). The spring ($\mbox{s=800.0 N/m}$) has linear characteristics with \MK{stretched length  $\mbox{a=0.25 m}$ and tensionless length $\mbox{l=0.17 m}$}. The damping in the system is given by a linear damper ($\mbox{c=10.0 Ns/m}$). As we consider discrete-time systems in the proposed input design approach, we discretize the system using the forward Euler method: 
\begin{equation}\label{eq:Euler}
    x_{k+1}=x_k+T_s \cdot f(x_k,u_k),
\end{equation}
\MS{where the function $f(x_k,u_k)$ is defined as:
\begin{equation}\label{eq:f(x,u)}
    f(x_k,u_k) = \begin{bmatrix}
            x_{2,k} \\
            \frac{1}{m} \REV{\left(F_k - sx_{1,k} + \frac{slx_{1,k} }{\sqrt{x_{1,k}^2 + a^2}} - cx_{2,k}\right)}
        \end{bmatrix}
\end{equation}
where $x_k$ and $u_k$ represent the state and input at time step $k$ respectively and the sampling time $T_s=\frac{1}{fs} = 0.01$ s.}

\subsection{Input parametrization}\label{seq:input parametrixation}
We considered a periodic multisine excitation with $F=184$ excited frequencies between \MK{the range $f_\mathrm{min} = 1$ Hz and $f_\mathrm{max} = 10$ Hz to cover the resonance frequency of the system}, a sample frequency $f_s = 100$ Hz, and $N = 2048$ points per period during this example:
\begin{equation}\label{eq:multisine}
    \MK{u_k = \sum_{l=l_{min}}^{l_{max}} A \cdot \sin(2\pi l \frac{f_0}{f_s} k + \varphi_l)}.
\end{equation}
where \MK{$l_\mathrm{min}=21$ and $l_\mathrm{max} = 204$}. The amplitude of the multisine is chosen such that it is zero-mean and has a standard deviation of \REV{160} Newton, while we optimize the signal by changing the phases of each frequency component: \MK{$\theta = \left[\varphi_1 \: \varphi_2 \: \ldots \: \varphi_F \right]$.}

By fixing the amplitude of each frequency component in our signal we have essentially constrained the energy content that is present in our signal. Hence, we will optimize the phases of the multisine signal while keeping the amplitude spectrum of the signal constant. Note that this is a strong constraint, which would be problematic for many input design approaches, however, the results will show that such a constraint does not pose a problem to the proposed method. Of course, other input parametrizations, with different constraints, might result in a better space-covering experiment.

The steady-state response of the system to the periodic multisine excitation is considered during optimization. The phases are initialized randomly, and distributed uniformly over the interval \MK{$[0,\:2\pi)$}.

\subsection{Considered Domain and Cost Function}
The domain of interest is selected as a rectangular box in the 3-dimensional input-state space. \REV{The dimension of axis $u$ is between $[-400;400]$ and of $x_2$ is between $[-20;20]$ with a grid distance of $42.1053$ and $2.1053$.} Our system varies less along axis $x_1$, therefore the domain is limited between $[-2;2]$ with a grid distance equal to $0.2105$. To have sufficient overlap between the region of influence of neighboring grid points, the kernel width of the cost function in direction $u$, $x_1$ and $x_2$ is \MK{$\sigma^2 = \left[\REV{40^2},\: 0.2^2,\: 2^2 \right]$.}

\subsection{Results}
\begin{figure}[htb]
    \centering
    \includegraphics[width=0.95\columnwidth]{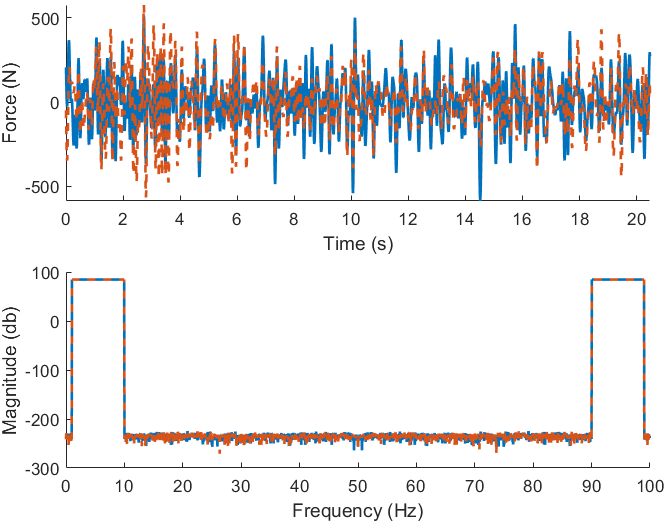}
    \caption{Initial (blue) and optimized (orange) input in time- and frequency domain.}
    \label{fig:Signal spectrum}
\end{figure}

Figure~\ref{fig:Signal spectrum} shows how the time-domain behavior of the signal changed by the optimization, \MK{while the amplitude spectrum remained fixed.} The cost \eqref{eq:overall cost} of the optimized signal is 69 times lower than the cost of the initial random phase multisine \REV{(3.88 vs 287.15)}. 

\begin{figure}[htb]
    \centering
    \includegraphics[width=0.95\columnwidth]{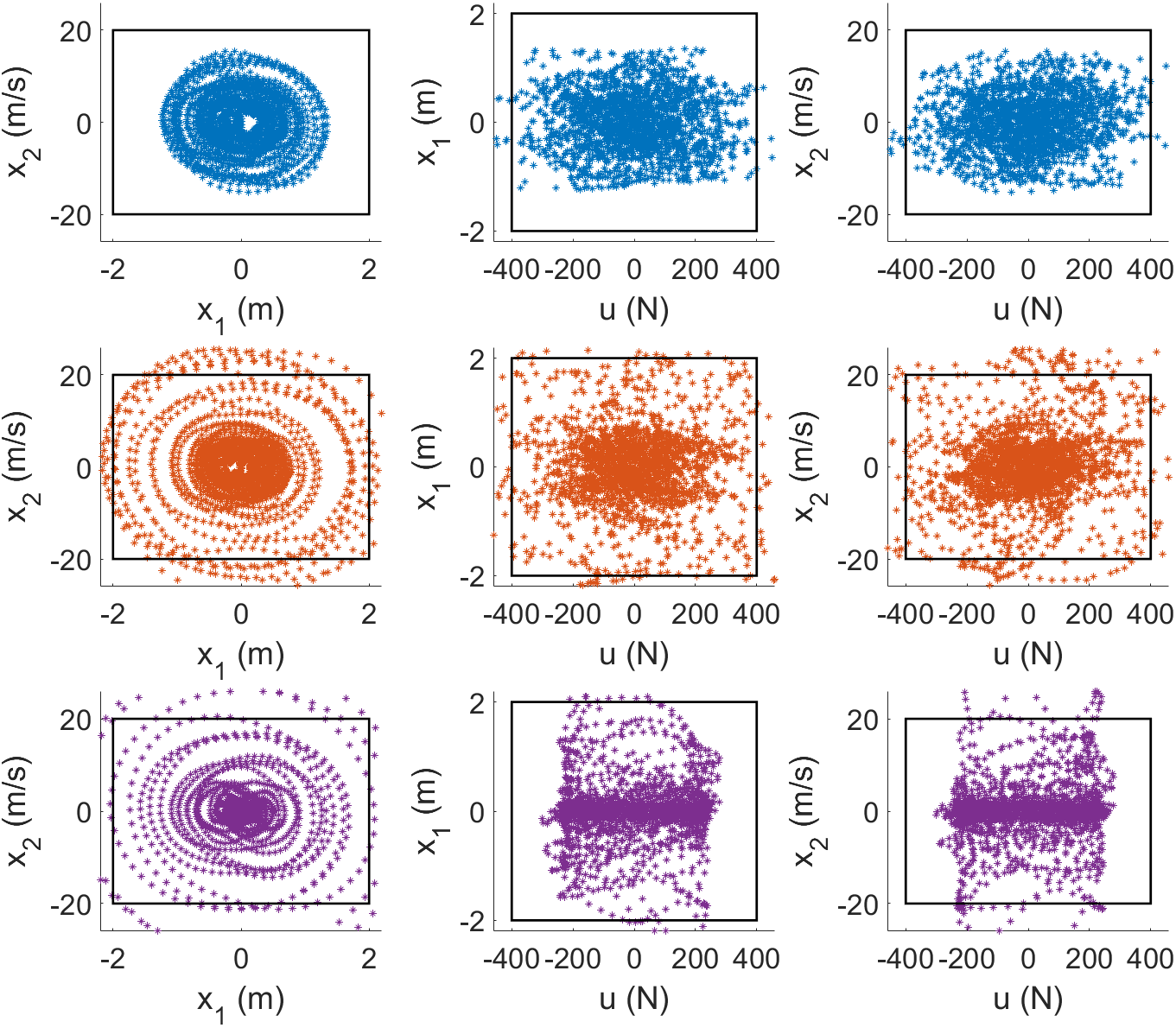}
    \caption{Side view \MS{of the space-filling effect of} the input before (blue) and after (orange) optimization \MS{and space-filling effect of the Schroeder multisine (purple)}. The black contour lines outline the borders of the domain of interest.}
    \label{fig:Side views}
\end{figure}

\begin{figure}[htb]
    \centering
    \includegraphics[width=0.95\columnwidth]{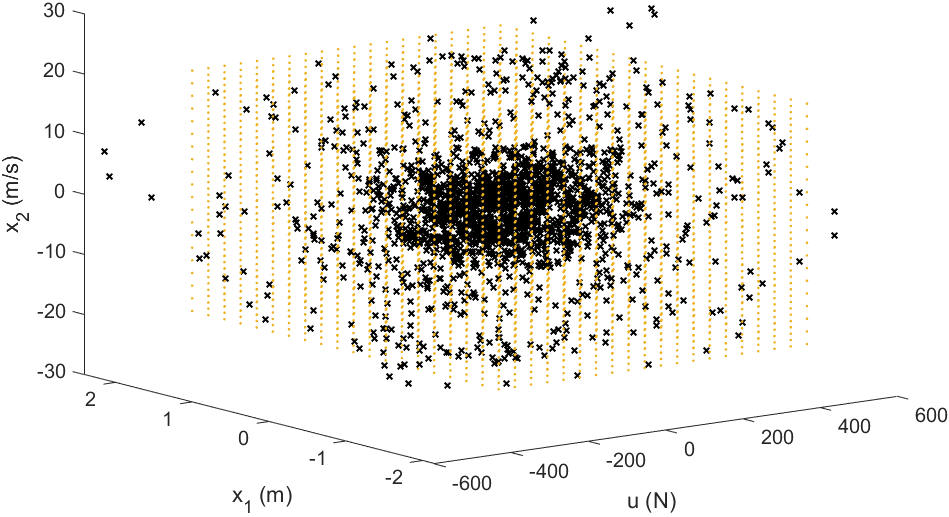}
    \caption{Optimized points in the 3D space. The grid points are shown by the yellow dots. The optimized input-state samples are shown by the black crosses.}
    \label{fig:3D plot}
\end{figure}

\begin{figure}[htb]
    \centering
    \includegraphics[width=0.95\columnwidth]{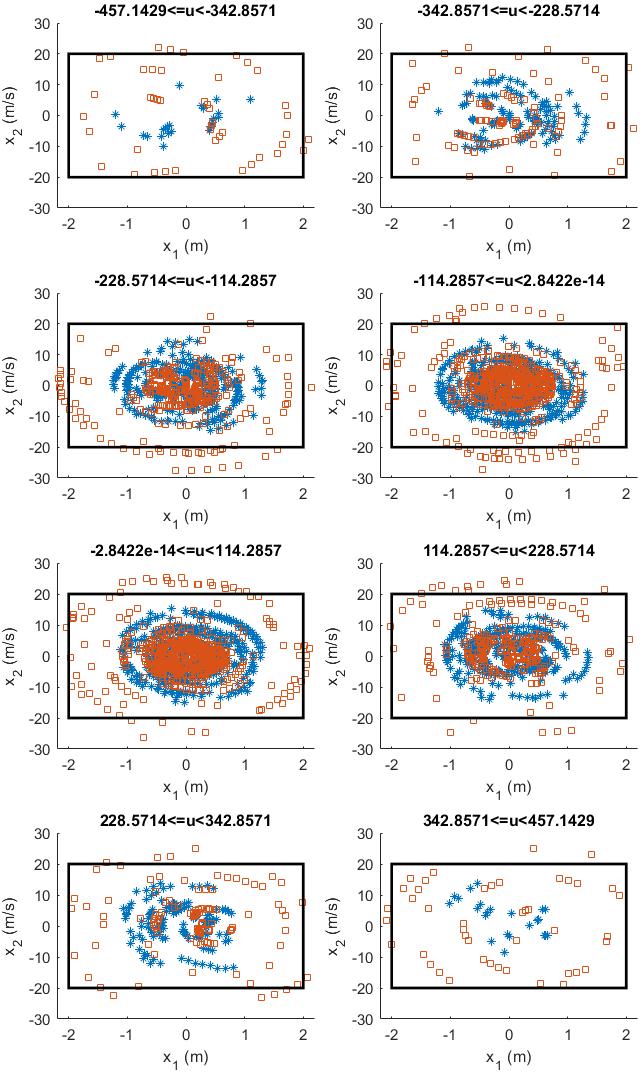}
    \caption{Cross-section view of the space-filling input before (blue) and after (orange) optimization sliced along the $u$ axis.}
    \label{fig:Cross-section view}
\end{figure}

Figure~\ref{fig:Side views} shows the side views of the considered 3-dimensional domain $\mathcal{D}_z$, a 3D depiction over the considered domain can be observed in Figure~\ref{fig:3D plot}, and sliced crosssections of the domain are shown in Figure~\ref{fig:Cross-section view}. While the initial input is strongly clustered around the domain center, a clear space-filling effect can be observed after optimization, this is best visible in the $x_1u$-plane. Nevertheless, due to the nature of the nonlinear dynamical system and the selected input parametrization, most data points are still clustered close to the center. This is especially visible at the crossections for the extreme input amplitudes: only few data points are scattered in these crosssections, though the space-filling optimized signal is visibly better spread than the initial random phase multisine. We would like to stress that these results are obtained while exciting only a limited range of frequencies and by using a signal of only $N=2048$ points long. This emphasizes the promising nature of the proposed approach. Note as well that the $x_1x_2$-plane shows a strong coherence between the two states, as can be expected from a dynamical system such as the mass-spring-damper system considered in this simulation example.



\subsection{\MK{Copmarison to Schroeder multisine}}
\begin{figure}[htb]
    \centering
    \includegraphics[width=0.95\columnwidth]{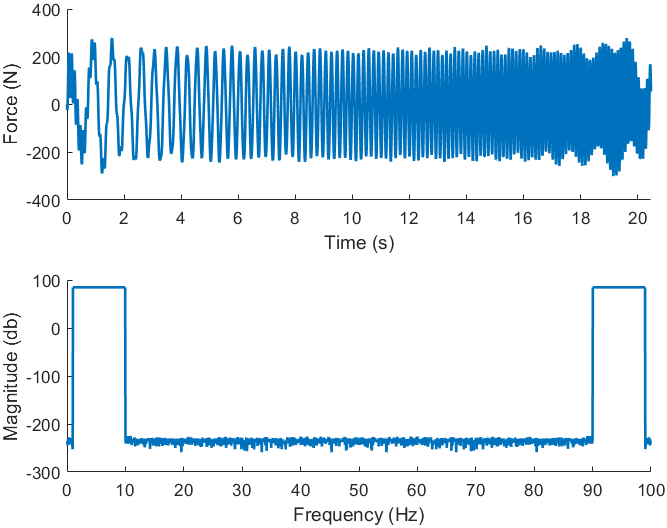}
    \caption{\MK{Multisine input signal with Schroeder phases in time- and frequency domain.}}
    \label{fig:Schroeder multisine}
\end{figure}

\MK{Furthermore, we compared our optimal solution to a Schroeder multisine. We constructed the new signal (Figure~\ref{fig:Schroeder multisine}) from (\ref{eq:multisine}) with the same parameters introduced in Section~\ref{seq:input parametrixation} with the exception of the phases. Each phase was constructed according to Schroeder`s equation:}
\begin{align}
    \MK{\varphi_l = \frac{-l(l-1) \pi}{N_l}} ,
\end{align}
\MK{where $N_l$ is the number of excited frequencies.}

\MK{Figure~\ref{fig:Side views} shows that indeed, in this case, the Schroeder multisine yields a better space-filling effect compared to the non-optimized random phase multisine, regardless of the data is also strongly clustered in the middle region. However, despite having the same amplitude spectrum, compared to the optimized input signal in this paper, the Schroeder phase fails to stretch out the data along the input axis ($u$). The crest factor in the case of the Schroeder multisine was 1.87 and in the case of the optimized signal was \REV{3.67}.}


\subsection{Monte-Carlo Simulation}\label{sec:MonteCarlo}
\begin{figure}[htb]
    \centering
    \includegraphics[width=0.85\columnwidth]{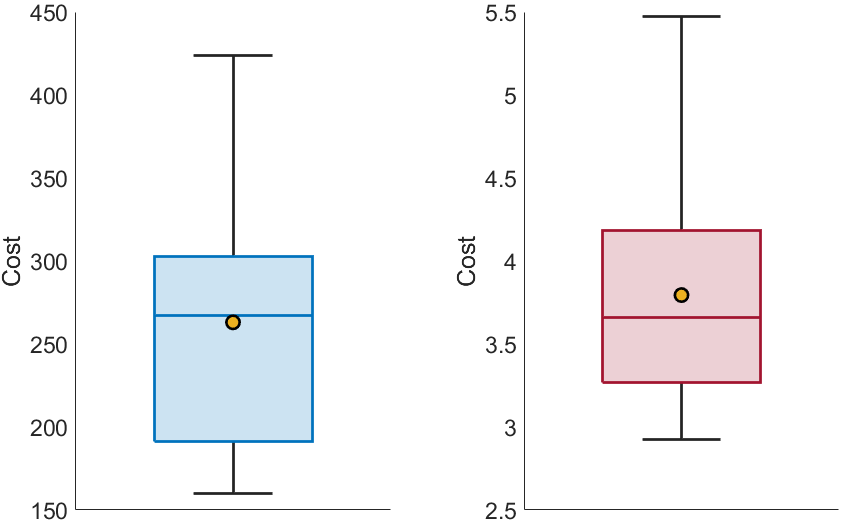}
    \caption{Box plot of the Monte-Carlo simulations of the costs before optimization (blue) and after optimization (red) with the mean of the starting and optimized costs (yellow circle). Note that the box plots are displayed on a different y-scale.}
    \label{fig:Box plot}
\end{figure}

In order to check how the quality varies with different random initializations, we conducted 20 Monte-Carlo simulations. In each simulation, we performed the optimization starting from a new random phase multisine while the amplitude constraint and optimization method remained unchanged. Figure~\ref{fig:Box plot} presents the outcome of the simulations, where the left figure (blue) represents the initial costs and the right figure represents the costs at the end of the 20 simulations. Even if the resulting signals are different, they are similar in terms of space-filling costs thus it can be stated that the proposed space-filling input design approach exhibits robust sensitivity against a random initialization.

\section{Conclusion}\label{sec:Conclusion}
This paper introduced a solution to the space-filling input design problem for state-space systems. The proposed solution allows for a wide range of parametrizations of the input and applies to a wide range of nonlinear dynamical systems. The main ingredients are \MK{based} on a variance argument when using nonparametric function regression, but are significantly modified using gridding and a squared-exponential membership function to make it computationally feasible. However, the resulting cost function remains non-convex and is minimized using gradient-based optimization approaches. The practical effectiveness of the approach is illustrated on a nonlinear mass-spring-damper system, using a multisine input signal.

\bibliography{ifacconf}             

\begin{thebibliography}{17}
\providecommand{\natexlab}[1]{#1}
\providecommand{\url}[1]{\texttt{#1}}
\providecommand{\urlprefix}{URL }
\expandafter\ifx\csname urlstyle\endcsname\relax
  \providecommand{\doi}[1]{doi:\discretionary{}{}{}#1}\else
  \providecommand{\doi}{doi:\discretionary{}{}{}\begingroup \urlstyle{rm}\Url}\fi

\bibitem[{Annergren et~al.(2017)Annergren, Larsson, Hjalmarsson, Bombois, and Wahlberg}]{Annergren2017}
Annergren, M., Larsson, C.A., Hjalmarsson, H., Bombois, X., and Wahlberg, B. (2017).
\newblock Application-oriented input design in system identification: Optimal input design for control.
\newblock \emph{IEEE Control Systems Magazine}, 37(2), 31--56.

\bibitem[{Beintema et~al.(2023)Beintema, Schoukens, and T{\'o}th}]{Beintema2023}
Beintema, G.I., Schoukens, M., and T{\'o}th, R. (2023).
\newblock Deep subspace encoders for nonlinear system identification.
\newblock \emph{Automatica}, 156.

\bibitem[{Bombois et~al.(2021)Bombois, Morelli, Hjalmarsson, Bako, and Colin}]{Bombois2021}
Bombois, X., Morelli, F., Hjalmarsson, H., Bako, L., and Colin, K. (2021).
\newblock Robust optimal identification experiment design for multisine excitation.
\newblock \emph{Automatica}, 125.

\bibitem[{Capone et~al.(2020)Capone, Noske, Umlauft, Beckers, Lederer, and Hirche}]{Capone20}
Capone, A., Noske, G., Umlauft, J., Beckers, T., Lederer, A., and Hirche, S. (2020).
\newblock Localized active learning of gaussian process state space models.
\newblock \emph{Proc of the Annual Conference on Learning for Dynamics and Control}, 490--499.

\bibitem[{Colin et~al.(2020)Colin, Bombois, Bako, and Morelli}]{Colin20}
Colin, K., Bombois, X., Bako, L., and Morelli, F. (2020).
\newblock Data informativity for the identification of particular parallel hammerstein systems.
\newblock \emph{21st IFAC World Congress}, 53, 1102--1107.

\bibitem[{De~Cock(2017)}]{Cock_Phd}
De~Cock, A. (2017).
\newblock \emph{D-Optimal Input Design for the Identification of Structured Nonlinear Systems}.
\newblock Ph.D. thesis, Vrije Universiteit Brussel, Brussels, Belgium.

\bibitem[{De~Cock et~al.(2016)De~Cock, Gevers, and Schoukens}]{Cock_D_optimal}
De~Cock, A., Gevers, M., and Schoukens, J. (2016).
\newblock D-optimal input design for nonlinear {FIR}-type systems: A dispersion-based approach.
\newblock \emph{Automatica}, 73, 88--100.

\bibitem[{Forgione et~al.(2014)Forgione, Bombois, Van~den Hof, and Hjalmarsson}]{Forgione2014}
Forgione, M., Bombois, X., Van~den Hof, P.M., and Hjalmarsson, H. (2014).
\newblock Experiment design for parameter estimation in nonlinear systems based on multilevel excitation.
\newblock In \emph{Proc. of the European Control Conference (ECC)}, 25--30.

\bibitem[{Heinz and Nelles(2017)}]{Nelles2017}
Heinz, T. and Nelles, O. (2017).
\newblock Iterative excitation signal design for nonlinear dynamic black-box models.
\newblock \emph{In Proc. of the International Conference on Knowledge Based and Intelligent Information and Engineering Systems}, 1054--1061.

\bibitem[{Hjalmarsson and Mårtensson(2007)}]{Hjalmarsson2007}
Hjalmarsson, H. and Mårtensson, J. (2007).
\newblock Optimal input design for identification of non-linear systems: Learning from the linear case.
\newblock \emph{In Proc. of the American Control Conference}.

\bibitem[{Nadaraya(1964)}]{Nadaraya1964}
Nadaraya, E.A. (1964).
\newblock On estimating regression.
\newblock \emph{Theory of Probability \& Its Applications}, 9(1), 141--142.

\bibitem[{Novara(2007)}]{Novarra2007}
Novara, C. (2007).
\newblock Experiment design in nonlinear set membership identification.
\newblock In \emph{Proc. of the 2007 American Control Conference}, 1566--1571.

\bibitem[{Paduart et~al.(2010)Paduart, Lauwers, Swevers, Smolders, Schoukens, and Pintelon}]{Paduart2010}
Paduart, J., Lauwers, L., Swevers, J., Smolders, K., Schoukens, J., and Pintelon, R. (2010).
\newblock Identification of nonlinear systems using polynomial nonlinear state space models.
\newblock \emph{Automatica}, 46(4), 647--656.

\bibitem[{Schoukens and Ljung(2019)}]{Schoukens19}
Schoukens, J. and Ljung, L. (2019).
\newblock Nonlinear system identification: A user-oriented road map.
\newblock \emph{IEEE Control Systems Magazine}, 39, 28--99.

\bibitem[{Smits and Nelles(2021)}]{Nelles21}
Smits, V. and Nelles, O. (2021).
\newblock Genetic optimization of excitation signals for nonlinear dynamic system identification.
\newblock \emph{In Proceedings of the 18th International Conference on Informatics in Control, Automation and Robotics}, 138--145.

\bibitem[{Smits and Nelles(2022)}]{Nelles22}
Smits, V. and Nelles, O. (2022).
\newblock Space-filling optimization of excitation signals for nonlinear system identification.
\newblock \emph{In Proc. of the 19th International Conference on Informatics in Control, Automation and Robotics}, 255--262.

\bibitem[{Vincent et~al.(2010)Vincent, Novara, Hsu, and Poolla}]{Vincent2010}
Vincent, T.L., Novara, C., Hsu, K., and Poolla, K. (2010).
\newblock Input design for structured nonlinear system identification.
\newblock \emph{Automatica}, 46(6), 990--998.

\end{thebibliography}
                                                   
\end{document}